\documentclass[preprint]{aa}
\usepackage{graphicx}
\usepackage{aas_macros}
\usepackage{natbib}
\usepackage{amsmath}
\usepackage{amssymb}
\usepackage{setspace}
\usepackage[utf8x]{inputenc}
\usepackage{latexsym}
\usepackage{color}
\usepackage{dcolumn}
\usepackage{bm}

\bibpunct{(}{)}{;}{a}{}{,}

\begin{document}
\title{Populations III.1 and  III.2 gamma-ray bursts:\\
Constraints on the  event rate for future radio and X-ray surveys} 

\author{R. S. de Souza\inst{1,2} \and
	N. Yoshida\inst{1} \and K. Ioka\inst{3}}
	
\offprints{Rafael S. de Souza \email{rafael.desouza@ipmu.jp}}

 \institute{$^{1}$IAG, Universidade de S\~{a}o Paulo, Rua do Mat\~{a}o 1226, Cidade
Universit\'{a}ria, CEP 05508-900, S\~{a}o Paulo, SP, Brazil\\
$^{2}$Institute for the Physics and Mathematics of the Universe, 
Todai Institutes for Advanced Study, 
University of Tokyo, 5-1-5 Kashiwanoha, Kashiwa, Chiba 277-8583, Japan\\
$^{3}$KEK Theory Center and the Graduate University for Advanced Studies (Sokendai),  Tsukuba 305-0801, Japan}

\date{Accepted -- Received}

\date{Released Xxxxx XX}

\authorrunning{de Souza, Yoshida \& Ioka}
   \titlerunning{ Population III.1 and  III.2 
Gamma-Ray Bursts}

\abstract
{}
{We calculate the theoretical event rate of gamma-ray bursts (GRBs)
from the collapse of massive first-generation (Population III; Pop III) stars.
The Pop III GRBs could be super-energetic
with the isotropic energy up to $E_{\rm iso} \gtrsim 10^{55-57}$ ergs,
providing a unique probe of the high-redshift Universe.}
{We consider both the so-called Pop III.1 stars (primordial)
and Pop III.2 stars (primordial but affected by radiation from other stars).
We employ a semi-analytical approach
that considers  inhomogeneous hydrogen reionization and
chemical evolution of the intergalactic medium. } 
{We show that Pop III.2 GRBs occur more than 100 times more
 frequently than Pop III.1 GRBs, and thus should be
 suitable targets for future GRB missions.
Interestingly,  our optimistic model predicts an event rate 
that is already constrained by the current radio transient searches.
We expect $\sim10-10^4$ radio afterglows above $\sim$ 0.3 mJy on the sky  
with $\sim 1$ year variability and mostly without GRBs (orphans), 
 which are detectable by ALMA, EVLA, LOFAR,  and SKA,
while we expect to observe maximum of $N < 20$ GRBs per year integrated over at $z>6$ for Pop III.2 and $N < 0.08$ per year integrated over at $z>10$ for Pop III.1 with EXIST, and $N < 0.2$ for Pop III.2 GRBs per year integrated over at $z > 6$ with \textit{Swift}.}
{}

\maketitle
\section{Introduction}
The first stars in the Universe are thought to have 
played a crucial role in the early cosmic evolution  
by emitting the first light and producing the first heavy elements
\citep{bromm09}. 
Theoretical studies based on the standard cosmological model
predict that the first stars were formed when 
the age of the Universe was less than a few hundred million years
and that they were predominantly massive
\citep{abel2002,omukai2003,yoshida2006}. Although there are
a few observational ways to probe the early stellar populations, 
there has not yet been any direct observations of the so-called
Population III (Pop~III) stars.

Recently, it has been proposed that massive Pop~III stars may produce 
collapsar gamma-ray bursts (GRBs), whose total isotropic energy 
could be $E_{\rm iso} \gtrsim 10^{55-57}$ ergs, i.e., $\gtrsim 2$ orders of magnitude above average
\citep{komissarov2010,toma2010,meszaros2010,barkov2010,suwa2011}. 
Even if the Pop III star has a supergiant
hydrogen envelope, 
the GRB jet can break out the first star because of the longlasting 
accretion of the envelope itself \citep{suwa2011,nagakura2011}. 
The observations of the spectral peak energy-peak luminosity relation also imply that the
 GRB formation rate does not significantly decrease to $z\sim 12$ \citep{Yonetoku2004, nava2011}.

Observations of such energetic GRBs at very high redshifts 
will provide a unique probe of the high-redshift Universe 
\citep{lamb2000,ciardi2000,Gou2004,Yonetoku2004}.  
\citet{ioka2005} study the radio afterglows of high-redshift GRBs 
and show that they can be observed out to $z\sim 30$ 
by VLA\footnote{http://www.vla.nrao.edu/}, LOFAR,\footnote{
http://www.lofar.org/index.htm} and
SKA\footnote{http://www.skatelescope.org/} \citep[see also][]{ioka2003}. 
\citet{inoue2007} calculate time-dependent, broad-band afterglow 
spectra of high-redshift GRBs.  They suggest that spectroscopic 
measurements of molecular and atomic absorption lines due to ambient 
protostellar gas may be possible to $z \sim 30$ and beyond 
with ALMA\footnote{www.alma.nrao.edu/}, 
EVLA\footnote{http://www.aoc.nrao.edu/evla/}, and SKA.
In the future, it will  be also promising  
to observe the GRB afterglows  located by gamma-ray satellites such as \textit{Swift}\footnote{http://swift.gsfc.nasa.gov/docs/swift/swiftsc.html},  SVOM\footnote{http://www.svom.fr/svom.html}, JANUS,\footnote{http://sms.msfc.nasa.gov/xenia/pdf/CCE2010/Burrows.pdf}
and EXIST\footnote{http://exist.gsfc.nasa.gov/design/}. 
Clearly, it is important to study the rate and the 
detectability of Pop~III GRBs at very high redshifts.

There have been already a few observations of GRBs at high redshifts.  
GRB 090429B at  $z = 9.4$ \citep{cucchiara2011} is the current 
record-holding object, followed by a $z = 8.6$ galaxy \citep{Lehnert2010}, 
GRB 090423 at  $z = 8.26$ 
\citep{Salvaterra2009,Tanvir2009}, GRB 080913 
at $z = 6.7$ \citep{Greiner2009},  GRB 050904 at $z = 6.3$ 
\citep{Kawai2006,Totani2006},  and the  highest redshift quasars 
at $z = 7.085$ \citep{mortlock2011} and $z = 6.41$ \citep{Willott2003}.  \citet{Chandra2010} report 
the discovery of radio afterglow emission from GRB 090423 and 
\citet{Frail2006} for GRB 050904. Observations of afterglows  
make it possible to derive the  physical properties of the explosion 
and the circumburst medium. It is intriguing to search for these different 
signatures in the GRB afterglows at low and high redshifts.

The purpose of the present paper is to calculate the Pop~III GRB rate  
detectable by the current and future GRB missions  \citep[see also][]{Campisi2011}.     
We consider high-redshift GRBs of two populations following \citet{bromm09}.
Pop~III.1 stars are the first-generation stars that 
form from initial conditions determined cosmologically.
Pop~III.2 stars are zero-metallicity stars but formed
from a primordial gas that was influenced by earlier
generation of stars. Typically, Pop~III.2 stars are formed
in an initially ionized gas \citep{Johnson06,Yoshida07}. 
The Pop~III.2 stars are thought to be 
less massive ($\sim 40$--$60 M_{\odot}$) than Pop~III.1 stars
($\sim 1000 M_{\odot}$)
but still massive enough for producing GRBs.

We have calculated the GRB rate for these two populations separately 
for the first time. The rest of the paper is organized as follows.
In Sect. 2, we describe a semi-analytical model to calculate 
the formation rate of primordial GRBs.  
In Sect. 3, we show our model predictions and 
calculate the detectability of Pop~III GRBs by
future satellite missions and by radio observations.
In Sect. 4, we discuss the results and give our concluding remarks.
Throughout the paper we adopt the standard $\Lambda$ cold dark matter 
model with the best fit cosmological parameters 
from   \citet{jarosik2010} 
(WMAP-Yr7\footnote{http://lambda.gsfc.nasa.gov/product/map/current/}),  
$\Omega_{\rm m} = 0.267, \Omega_{\Lambda} = 0.734$,  
and $H_0 = 71 {\rm km}~{\rm s}^{-1}{\rm Mpc}^{-1}$.

\section{Gamma-ray burst  rate}

We assume that the formation rate of GRBs is proportional 
to the star formation rate \citep{totani1997,ishida2011}. 
The number of observable GRBs per comoving volume per time is expressed as
\begin{equation}
\Psi_{\rm GRB}^{obs}(z) = \frac{\Omega_{obs}}{4\pi}\eta_{\rm GRB}\,\eta_{\rm beam}\,
\Psi_{*}(z)\int_{\log{L_{\rm lim}(z)}}^{\infty}p(L)d\log{L},
\label{psigrb}
\end{equation}
where $\eta_{\rm GRB}$ is the GRB formation efficiency (see section 2.6),  
$\eta_{\rm beam}$  the beaming factor of the burst,  $\Omega_{obs}$  the field of view of the experiment, 
$\Psi_{*}$  the cosmic star formation rate (SFR) density,  
and $p(L)$  the GRB luminosity function in X-rays to gamma rays. The intrinsic GRB rate is given by
\begin{equation}
 \Psi_{\rm GRB}(z) = \eta_{\rm GRB}\Psi_{*}(z).
 \label{psigrbreal}
 \end{equation}
The quantity $L_{\rm lim}(z)$ is the minimum 
luminosity threshold to be detected, which is specified 
for a given experiment.   
The nonisotropic nature of GRBs gives
$\eta_{\rm beam} \sim 0.01-0.02$  \citep{guetta2005}.  Using a radio transient survey,    
\citet{gal2006} place an upper limit of $\eta_{\rm beam} \lesssim 0.016$.
Given the average value of jet opening angle $\theta \sim 6^\circ$ 
\citep{ghirlanda2007} and $\eta_{\rm beam} \sim 5.5\times10^{-3}$, 
we set $\eta_{\rm beam} = 0.006$ as a fiducial value.
The adopted values of $\Omega_{\rm obs}$ are 1.4, 2, 4,  and 5 
for \textit{Swift}, SVOM, JANUS,  and EXIST,  respectively \citep{Salvaterra2008}.  
 
\subsection{The number of collapsed objects}

We first calculate the star formation rate (SFR) at early epochs.  
Assuming that stars are formed in collapsed dark matter halos,   
we follow a popular prescription in which the number of collapsed
objects is calculated by the halo mass function 
\citep{Hernquist2003,greif2006,trenti2009}.  
We adopt the Sheth-Tormen mass function, $f_{\rm ST}$, \citep{sheth1999} 
to estimate  the number of dark matter halos, $n_{\rm ST}(M,z)$, 
with mass less than  $M$  per comoving volume at a given redshift:

\begin{equation}
f_{\rm ST} = A\sqrt{\frac{2a_1}{\pi}}
\left[1+\left(\frac{\sigma^2}{a_1\delta_{c}^2}\right)^p\right]
\frac{\delta_{c}}{\sigma}\exp{\left[-\frac{a_1\delta_{c}^2}{2\sigma^2}\right]}, 
\end{equation}
where $A=0.3222, a_1 = 0.707, p = 0.3$ and  $\delta_c = 1.686$. The mass function $f_{\rm ST}$ can be related to the  $n_{\rm ST}(M,z)$ as
\begin{equation}
f_{\rm ST} = \frac{M}{\rho_{\rm m}}\frac{dn_{\rm ST}(M,z)}{d\ln{\sigma^{-1}}}, 
\end{equation}
 where $\rho_{\rm {m}}$ is the total mass density of the background Universe. 
The variance of the linear density field $\sigma (M,z)$ is given by 
\begin{equation}
\sigma^{2}(M,z) = \frac{b^2(z)}{2\pi^2}\int_{0}^{\infty} k^2P(k)W^2(k,M)dk,
\end{equation}
where $b(z)$ is the growth factor of linear perturbations normalized 
to $b = 1$ at the present epoch, and $W(k,M)$ is the Fourier-space top hat filter. 
To calculate the power spectrum $P(k)$, we use the CAMB code\footnote{http://camb.info/} 
for our assumed $\Lambda$CDM cosmology. 

\subsection{$\rm H_2$ Photodissociation}

The star formation efficiency in the early Universe largely
depends on the ability of a primordial gas to cool and condense.
Hydrogen molecules (H$_{2}$) are the primary coolant
in a gas in small mass ``minihalos'', and 
are also fragile to soft ultraviolet radiation,
and thus a ultraviolet background in the Lyman-Werner (LW) bands 
can easily suppress the star formation inside minihalos. 
We model the dissociation effect by setting the minimum mass for halos that 
are able to host Pop~III stars \citep{yoshida2003}.

For the minimum halo mass capable of cooling by molecular hydrogen
in the presence of a Lyman-Werner (LW) background, we adopt a
fitting formula given by \citet{Machacek2001} and \citet{wise2005}, which  also 
agrees with results from \citet{O'Shea2008}: 
\begin{equation}
M_{\rm H_2} = \exp{\left(\frac{f_{\rm cd}}{0.06}\right)}
(1.25 \times 10^{5} + 8.7\times10^{5}F_{\rm LW,-21}^{0.47}), 
\label{eq:minmass}
\end{equation}
where $F_{\rm LW,-21}= 4\pi J_{\rm LW}$ is the flux in the LW band in units 
of $10^{-21} {\rm erg}^{-1} {\rm s}^{-1} {\rm cm}^{-2} {\rm Hz}^{-1}$,  and
$f_{\rm cd}$  the fraction of gas that is cold and dense.  
We set $f_{\rm cd} = 0.02$ as a conservative estimate.
We compute the LW flux consistently with the comoving density in stars $\rho_{*}(z)$
via a conversion efficiency $\eta_{\rm LW}$ \citep{greif2006}: 
\begin{equation}
J_{\rm LW} = \frac{hc}{4\pi m_{\rm H}}\eta_{\rm LW}\rho_{*}(z)(1+z)^3.
\end{equation}
Here, $\eta_{\rm LW}$ is the number of photons emitted in the LW bands 
per stellar baryon and $m_{\rm H}$  is the mass of hydrogen. 
The value of $\eta_{\rm LW}$ depends on the characteristic mass
of the formed primordial stars,
but the variation is not very large
for stars with masses greater than ten solar masses \citep{schaerer2002}. 
We set  $\eta_{\rm LW} = 10^4$ for both Pop~III.1 and Pop~III.2 for
simplicity.

Next we calculate the stellar mass density as
\begin{equation}
\rho_{*}(z) = \int \Psi_{*}(z')\left|\frac{dt}{dz^{'}}\right|dz'.
\end{equation}
For a given \emph{z}, the integral is performed over the maximum distance 
that an LW photon can travel before it is redshifted out of the LW bands. 
The mean free path of LW photons at z = 30 is $\sim 10$ Mpc (physical).  
Photons travel over the mean free path in $\sim 10^7$ yr \citep{Mackey2003}. 
Halos with virial temperature less than $10^4$ Kelvin cool 
almost exclusively by H$_2$ line cooling, 
and produce mostly massive stars. 
We adopt the mass of such halos, $M (T_{\rm vir} = 10^{4} {\rm K}, z)$, 
as an upper limit of halos that produce Pop~III.1 stars. 
In larger halos, the gas is ionized at virialization,
and thus the formed stars have, according to our definition,
similar properties to Pop~III.2 stars. 
We assume that $M (T_{\rm vir} = 10^{4} {\rm K}, z)$ 
is the minimum halo mass for Pop~III.2 star formation.

The collapsed fraction of mass, $F_{\rm {col}}(z)$, available for Pop~III star 
formation is given by
\begin{equation}
F_{\rm {col}}^{\rm III.1}(z)=\frac{1}{\rho_{\rm {m}}}\int_{M_{\rm H_2}}^{M_{T_{\rm vir} = 10^{4} {\rm K}}} dMMn_{\rm {ST}}(M,z)
\label{fcol}
\end{equation}
for Pop III.1 stars, and 
\begin{equation}
F_{\rm {col}}^{\rm III.2}(z)=\frac{1}{\rho_{\rm {m}}}\int_{M_{T_{\rm vir} = 10^{4} {\rm K}}}^{\infty} dMMn_{\rm {ST}}(M,z)
\label{fcol2}
\end{equation}
for Pop III.2. 
Using the  above criteria, the SFR of Pop~III stars can be written as
\begin{equation}
\Psi_{*}^{\rm III.1}(z)=(1-Q_{\rm H II}(z))(1-\zeta(z,v_{\rm wind}))\rho_{\rm {m}}
f_b f_{*}\frac{\rm {d}F^{\rm III.1}_{\rm {col}}}{dt}
\label{SFRH}
\end{equation}
for Pop III.1 stars, and 
\begin{equation}
\Psi_{*}^{\rm III.2}(z)=Q_{\rm H II}(z)(1-\zeta(z,v_{\rm wind}))\rho_{\rm {m}}
f_b f_{*}\frac{\rm {d}F^{\rm III.2}_{\rm {col}}}{dt}
\label{SFRH2}
\end{equation}
for Pop III.2. 
Here, $\zeta(z,v_{wind})$ represents the global filling fraction of metals 
via galactic winds (see section 2.4),  $Q_{\rm H II}(z)$ the volume filling fraction of ionized
regions (see section 2.3),  and $f_b$ is the baryonic mass fraction.  
 For the star formation efficiency, we use the value 
$f_{*} = 0.001$  as a conservative choice \citep{greif2006} 
and $f_{*} = 0.1$ \citep{bromm2006} as an upper limit.  The latter
choice is not strictly consistent with the assumption made in Eq. (\ref{eq:minmass}).
We explore a model with $f_{*} = 0.1$ simply to show a very optimistic case.

\subsection{Reionization}
Inside growing {H\sc{ii}}  regions, the gas is highly ionized,  and 
the temperature is $\sim 10^4$ K, so the formation of Pop III.1 stars 
is terminated according to our definition. 
The formation rate of Pop III.1 is reduced
by a factor given by the volume filling fraction of ionized
regions, $Q_{\rm H II}(z)$.  
We follow \citet{wyithe2003} to calculate
the evolution of $Q_{\rm {H\sc{II}}}(z)$ as
\begin{equation}
\frac{dQ_{\rm H II}}{dz} 
= \frac{N_{\rm ion}}{0.76}\frac{dF_{\rm col}}{dt}-\alpha_{\rm B}\frac{C}{a^3}n_{\rm H}^0Q_{\rm H II}, 
\end{equation}
whose solution is 
\begin{equation}
Q_{\rm H II}(z) = 
\int_{z}^{\infty}dz'\frac{dt}{dz'}\frac{N_{\rm ion}}{0.76}\frac{dF_{\rm col}}{dt}e^{F(z',z)},
\end{equation}
where 
\begin{equation}
F(z',z) = -\frac{2}{3}\frac{\alpha_{\rm B} n_{\rm H}^0}{\sqrt{\Omega_{\rm m}}H_0}C[f(z')-f(z)],
\end{equation}
and 
\begin{equation}
f(z) = \sqrt{(1+z)^3+\frac{1-\Omega_{\rm m}}{\Omega_{\rm m}}}.
\end{equation}

Here we have assumed the primordial fraction of hydrogen of 0.76. 
In the above equations, $N_{\rm ion}\equiv N_{\gamma}f_{*}f_{\rm esc}$ is an efficiency 
parameter that gives the number of ionizing photons per baryon, where 
$f_{esc}$ is the fraction of ionizing photons able to escape the host galaxy,  and 
$N_{\gamma}$ is the time averaged number of ionizing photons emitted per unit 
stellar mass formed. 
The quantity $n_{\rm H}^0 = 1.95\times 10^{-7} {\rm cm}^{-3}$ is the present-day
comoving number  density of hydrogen,  
$\alpha_{\rm B} = 2.6\times 10^{-13} {\rm cm}^3 {\rm s}^{-1} $ is the hydrogen 
recombination rate,  and $C = \langle n_{\rm H}^2\rangle/\bar{n}_{\rm H}^2$  the clumping factor.    
We use the average value $C =4$ (see \citet{Pawlik2009} for detailed discussion 
about redshift dependence of C).
We set the values $f_{esc} = 0.7, f_{*} = 0.01$,  and  $N_{\gamma} = 9\times 10^{4}$ 
as fiducial values \citep{greif2006}. 

\begin{figure}
\includegraphics[width=0.9\columnwidth]{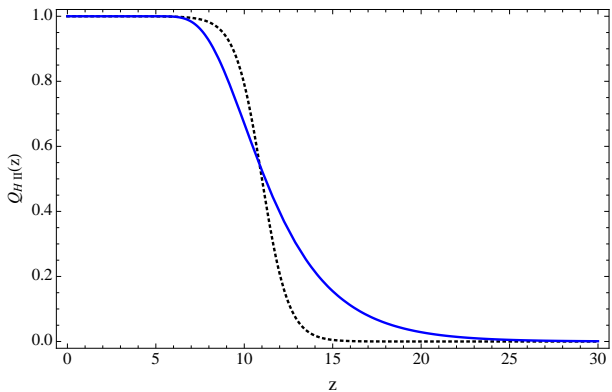}
\caption{Reionization history calculated using  our model. 
The blue line is our model prediction, and the dotted black
line is the best fit of 
CAMB code.}
\label{fig:QHII}
\end{figure}

 In Fig. \ref{fig:QHII} we show the reionization history calculated 
using our model in comparison with a fitting function that is the default 
parametrization of reionization in CAMB \citep{lewis2000}. 

\subsection{Metal enrichment}

We need to consider metal enrichment in the intergalactic medium (IGM)
in order to determine when the formation of primordial  
stars is terminated (locally) and when the star formation 
switches from the Pop~III mode to a more conventional one.  

It is thought that Pop~III stars do not generate strong stellar winds, 
and thus the main contribution 
to the metal pollution comes from their supernova explosions.  
\citet{madau2001} argue that pregalactic outflows from the same primordial halos 
that reionize the IGM could also pollute it with a substantial 
amount of heavy elements.
To incorporate the effect of metal enrichment by galactic winds,
we adopt a similar prescription to  \citet{jarret2010} and \citet{Furlanetto2005}. 
 
We assume that star-forming halos (``galaxies'') launch a wind of metal-enriched
gas at $z_{*}$ $\sim 20$. The metal-enriched wind propagates outward from 
a central galaxy with a velocity $v_{\rm wind}$, traveling over a comoving distance 
$R_{\rm wind}$ given by 
\begin{equation}
R_{\rm wind} = \int_{z_{*}}^{z}v_{\rm wind}(1+z')\frac{dt}{dz'}dz'.
\end{equation}
Then we can express $f_{chem}$, the ratio of gas mass enriched by the wind 
to the total gas mass in each halo, as
\begin{equation}
f_{\rm chem}(M,z,v_{\rm wind}) = \frac{4\pi}{3}\frac{R_{\rm wind}^3}{V_{\rm H}},
\label{eqchem}
\end{equation}
 where $V_{\rm H}\propto R_{\rm H}^3$ is the volume of each halo. 
The halo radius $R_{\rm H}$ can be approximated by 
 \begin{equation}
 R_{\rm H}(M) = \left(\frac{3M}{4\pi\times180\rho_{\rm m}}\right)^{1/3}.
 \end{equation} 
 Equation  (\ref{eqchem}) takes  the self-enrichment of each halo into account. 
The next step is to  evaluate the average metallicity over cosmic scales.
The fraction of cosmic volume enriched by the winds can then be written as
\begin{equation}
\zeta(z,v_{\rm wind}) = \frac{1}{\rho_{\rm {m}}}\int dM f_{\rm b}f_{*}
f_{\rm chem}(M,z,v_{\rm wind})M n_{\rm {ST}}(M,z).
\end{equation}
Although this may appear a significant oversimplification, the model 
with $v_{\rm wind}$ as a single parameter  indeed provides  good insight 
into the impact of the metal enrichment.

We adopt three different values of $v_{\rm wind}$ and examine the effect
of metal enrichment quantitatively.  We assume that Pop~III stars
are not formed in a metal-enriched region, regardless of the 
actual metallicity. Even a single pair instability supernova 
can enrich the gas within a small  halo to a metallicity level well above 
the critical  metallicity (see e.g. \citet{schneider2006}). 
We effectively assume that the so-called critical 
metallicity is very low \citep{schneider2002,schneider2003,bromm2003,omukai2005,frebel2007,Krzysztof2010}. 

\begin{figure}
\includegraphics[width=0.9\columnwidth]{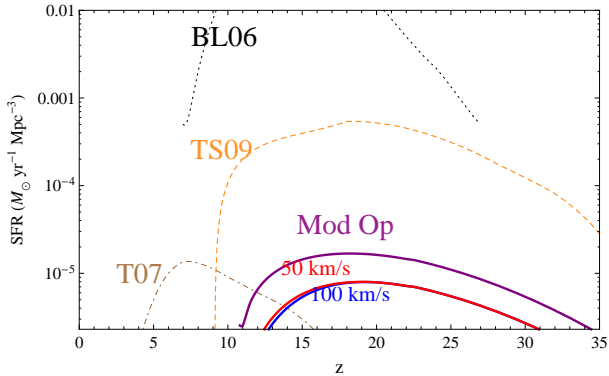}
\caption{Pop~III.1 star formation rate. Calculated for for weak and strong  chemical 
feedback models and a moderate star formation efficiency with $f_{*} = 0.05$.
The results are shown for $v_{\rm wind} = 50$ km/s , red line; 
and $100$ km/s, blue line. 
We also show the theoretical SFRs in the literature, 
from Bromm \& Loeb  2006 (Pop III.1+III.2), dotted  black line;
Trenti \& Stiavelli 2009 (Pop III.2), dashed orange line;
and Tornatore et al. 2007 (Pop III.1+III.2), dot-dashed brown line. 
The  purple line is our optimistic model
where we assume a very high star formation efficiency, $f_{*}\sim 0.1$,  and 
low chemical enrichment, $v_{\rm wind} =$ 50 km/s.}
\label{fig:SFR}
\end{figure}

\begin{figure}
\includegraphics[width=0.9\columnwidth]{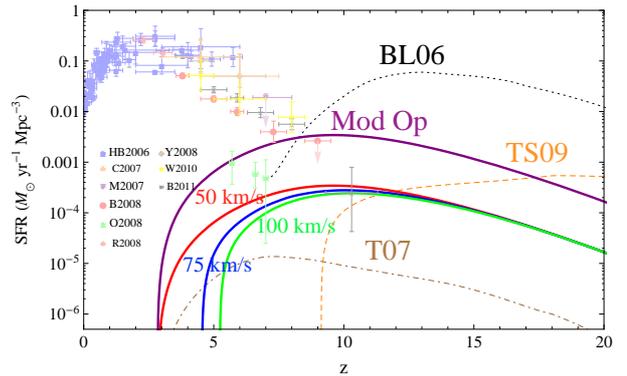}
\caption{Pop~III.2 star formation rate. Calculated  for three different chemical 
feedback models; 
$v_{\rm wind} = 50$ km/s,  red line;  
$v_{\rm wind} = 75$ km/s,  blue line; 
and $v_{\rm wind} = 100$ km/s,  green line.
We also show the theoretical SFRs in the literature, 
from Bromm \& Loeb  2006 (Pop III.1+III.2), dotted black line;
Trenti \& Stiavelli 2009 (Pop III.2), dashed orange line;
and Tornatore et al. 2007 (Pop III.1+III.2), dot-dashed brown line. 
The  purple line is our optimistic model
where we assume a very high star formation efficiency, $f_{*}\sim 0.01$,  and
low chemical enrichment, $v_{\rm wind} =$ 50 km/s. The light points are  independent SFR determinations compiled from the  literature. }
\label{fig:SFRII}
\end{figure}

Figures \ref{fig:SFR} and \ref{fig:SFRII} show the star formation rate  (SFR) history  
for both Pop~III.1 and Pop~III.2 considering three different 
values of the galactic wind, $v_{wind} = 50, 75, 100$ km/s. 
Figure \ref{fig:SFR} shows that the  metal enrichment has  
little   influence on Pop III.1.
This is because Pop III.1 formation is terminated early 
due reionization. In Fig. \ref{fig:SFRII} we compare the Pop~III.2 SFR history with a compilation of independent measures from \citet{hopkins2006} up to $z \approx 6$ 
and from observations of color-selected Lyman break galaxies \citep{mannucci2007, bouwens2008, bouwens2011},  Ly$\alpha$ Emitters \citep{ota2008}, UV+IR measurements \citep{reddy2008},  and GRB observations \citep{chary2007, yuksel2008, wang2009} at higher $z$ (hereafter, these will be referred to as H2006, M2007, B2008, B2011, O2008, R2008, C2007, Y2008,  and W2009, respectively).  The optimistic case for Pop III.2 is chosen to keep the SFR always below the observationally determined SFR 
at $z < 8$.

We compared our model results with the SFRs estimated by other authors 
in the literature \citet{bromm2006}, \citet{tornatore2007}, \citet{trenti2009}, and  also see \citet{naoz2007}.
It is important to note that Pop~III formation can continue to low 
redshifts ($z<10$) depending on the level of metal enrichment. 
\citet{tornatore2007} use cosmological simulations to show
that, because of limited efficiency of heavy element transport 
by outflows, Pop~III star formation continues to form down 
to \emph{z} = 2.5 (which intriguingly matches  our model with 
$v_{\rm wind}$ = 50 km/s in Fig. \ref{fig:SFRII}).  
The SFR of \citet{tornatore2007} 
has a peak value of $10^{-5} M_{\odot} {\rm yr}^{-1} {\rm Mpc}^{-3}$ 
at $z \approx 6$. 

\begin{figure}
\includegraphics[width=0.9\columnwidth]{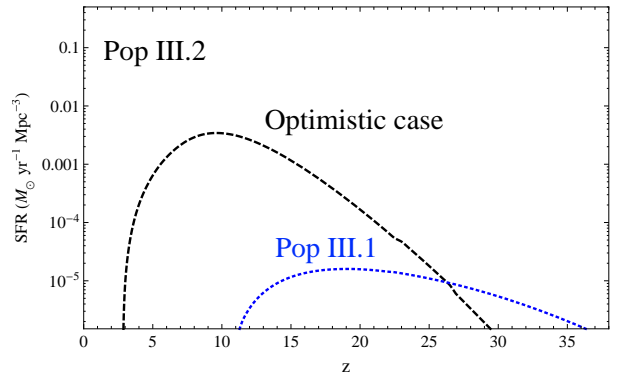}
\caption{ Comparison of  the star formation rates for Pop III.1 (blue dotted line) 
and for PopIII.2 (dashed black line), for our optimistic model with a high star formation efficiency $f_{*} = 0.1$ for Pop III.1,  $f_{*} = 0.01$ for Pop III.2 and slow chemical enrichment $v_{\rm wind} = 50 {\rm km/s}$.}
\label{fig:SFRIII_1III_2}
\end{figure}

In Fig. \ref{fig:SFRIII_1III_2} we also show the result of our model with $f_{*} = 0.1-0.01$ and
$v_{\rm wind} =$ 50 km/s for both Pop~III.1 and Pop~III.2,  respectively.
This model provides an ``optimistic'' estimate for the detectable 
GRB rate for the future missions (see Sect. 3). 

\subsection{Luminosity function}

The number of GRBs detectable by any given instrument depends on the 
instrument-specific flux sensitivity threshold and also on the intrinsic 
isotropic luminosity function of GRBs. For the latter,
we adopt the power-law distribution function  
of \citet{wanderman2010}
\begin{equation}
\label{LF}
p(L) =\left\{ \begin{array}{ll}\left(\frac{L}{L_{*}}\right)^{-0.2^{+0.2}_{-0.1}} &
L<L_{*}, \\ 
\left(\frac{L}{L_{*}}\right)^{-1.4^{+0.3}_{-0.6}} &
L >L_{*}. \\
 \end{array}\right.,
\end{equation}
where $L_{*}$ is the characteristic isotropic  luminosity. 
We set $L_{*}\sim 10^{53} ergs/s$ for Pop~III.1,   
whereas $L_{*}\sim 10^{52} ergs/s$ for Pop~III.2 stars
are similar to ordinary GRBs \citep{li2008, wanderman2010}. 
The Pop III.1 GRBs are assumed to be energetic with isotropic 
kinetic energy $E_{iso}\sim 10^{56-57} erg$ but long-lived $T_{90} \sim 1000~ s$, 
so that the luminosity would be moderate  
$L_{*}\sim\epsilon_{\gamma}\times 10^{56-57}/1000\sim10^{52-53}$ ergs/s 
if $\epsilon_{\gamma}\sim 0.1$ is the conversion efficiency from the jet kinetic 
energy to gamma rays  \citep{suwa2011}.

Using the above relation we can predict the observable GRB rate 
for the \textit{Swift},  SVOM, JANUS,  and EXIST missions.   
For \textit{Swift}, we set a bolometric energy flux limit 
$F_{\rm lim} = 1.2 \times 10^{-8} {\rm erg}~ {\rm cm}^{-2}~ {\rm s}^{-1}$ 
\citep{li2008}. We adopt a similar limit for SVOM \citep{paul2011}.  
For JANUS, $F_{\rm lim} \sim 10^{-8} {\rm erg}~ {\rm cm}^{-2}~ {\rm s}^{-1}$ 
\citep{janus2009}.   
The luminosity threshold is then
\begin{equation}
  L_{\rm lim} = 4\pi\, d_{\rm L}^{2}\, F_{\rm lim}.
\end{equation}
Here $d_L$ is the luminosity distance for the adopted $\Lambda$CDM cosmology. 
EXIST is expected to be $\sim 7-10$ times more sensitive 
than \textit{Swift} \citep{Grindlay2010}. 
We set the EXIST sensitivity threshold to ten times lower  
than \textit{Swift} as an approximate estimate. 
For simplicity,  we assume that the spectral energy distribution (SED) 
peaks at X-to-$\gamma$ ray energy (detector bandwidth) as an optimistic case.   
The rate would be less if the SED is very different from Pop II/I GRBs.
  
\subsection{Initial mass function and GRB formation efficiency}

The stellar initial mass function (IMF) is critically
important for determining  the Pop~III GRB rate.
We define the GRB formation efficiency factor per stellar mass as
\begin{equation}
\eta_{\rm GRB} =  f_{\rm GRB} \frac{\int_{M_{\rm GRB}}^{M_{\rm up}}\phi(m)dm}
{\int_{M_{\rm low}}^{M_{\rm up}}m\phi(m)dm}, 
\label{etagrb}
\end{equation}
where $\phi(m)$ is the stellar IMF,  and $f_{\rm GRB} = 0.001$
is the GRB fraction,  
since we expect one GRB every 1000 supernovae \citep{langer2006}.  
We assume that Pop III GRBs have a similar fraction to the standard case.  
\citet{izzard2004} argue that GRB formation efficiency
could increase by a factor of 5-7 for low-metallicity stars 
($\sim 10^{-2}Z_{\odot}$). If most of the first stars are rotating rapidly 
as suggested by \citet{stacy2010}, we can expect that a significant fraction
of the first stars 
can produce GRBs. Thus given the uncertainty  in the parameter for free metal 
stars,  we also explore the possibility that $f_{\rm GRB}$ is $10-100 \times$ higher.

We consider the following two forms of IMF.
One is a power law with the standard Salpeter slope
\begin{equation}  
\phi (m) \propto m^{-2.35},
\end{equation} 
and the other is a Gaussian 
IMF \citep{scannapieco2003,nakamura2001}:  
\begin{equation}
\phi(m)m^{-1}\, {\rm d}m = 
\frac{1}{\sqrt{2\pi}\sigma_{\rm c}(M)}
e^{-(m-\bar{M})^2/2\sigma_{\rm c}(M)^{2}} {\rm d} m.
\end{equation}
For the latter, we assume $\bar{M} = 550 M_{\odot}$ for Pop~III.1 
and $\bar{M} = 55 M_{\odot}$ for Pop~III.2, 
with dispersion $\sigma_c =  (\bar{M}-M_{\rm low})/3$.  
$M_{\rm low}$ is the minimum mass 
for a given stellar type,  $100 M_{\odot}$ for Pop~III.1,  and  
$10 M_{\odot}$ for Pop~III.2, whereas
$M_{\rm up}$ is the maximum mass for a given stellar type, 
$1000 M_{\odot}$ for Pop~III.1,  and $\sim 100 M_{\odot}$ for Pop~III.2.  
$M_{\rm GRB}$ is the minimum mass that is  able to trigger 
GRBs, which we set to be $25 M_{\odot}$ \citep{bromm2006}. 
Not all Pop~III.1 stars will leave a black hole behind 
at their deaths. In the narrow mass range of $\sim 140-260 M_{\odot}$ 
Pop~III stars are predicted to undergo a pair-instability supernova (PISN) 
explosion \citep{heger2002}. 
This range of mass is excluded from  the calculation of Eq. (\ref{etagrb}).

The efficiency factor for the power-law (Salpeter) IMF 
is $\eta_{\rm GRB}/f_{\rm GRB} \sim 1/926 M_{\odot}^{-1}$ and 
$1/87 M_{\odot}^{-1}$ for Pop~III.1 and Pop~III.2,   
respectively. Using the Gaussian IMF,  
$\eta_{\rm GRB}/f_{\rm GRB} \sim 1/538 M_{\odot}^{-1}$ and 
$1/53 M_{\odot}^{-1}$ for Pop~III.1 and Pop~III.2. respectively. 
Thus, the  GRB formation efficiency  for Pop III.2  can be about an 
order of magnitude higher than  Pop III.1  because of the lower 
characteristic mass of Pop III.2 stars.

\section{Redshift distribution of GRBs}
Over a particular time interval,  $\Delta t_{\rm obs}$,  in the observer rest frame, 
the number of observed GRBs originating between redshifts $z$ and $z + dz$
is
\begin{equation}
\frac{{\rm d}N_{\rm GRB}^{\rm obs}}{{\rm d}z} = \Psi_{\rm GRB}^{\rm obs}(z)\frac{\Delta t_{\rm obs}}{1+z}
\frac{{\rm d}V}{{\rm d}z}, 
\label{dngrbobs}
\end{equation}
where ${\rm d}V/{\rm d}z$ is the comoving volume element per unit redshift, given by
\begin{equation}
\frac{{\rm d}V}{{\rm d}z} = \frac{4\pi\, c\, d_{\rm L}^2}{(1+z)}\left|\frac{{\rm d}t}{{\rm d}z}\right|.
\end{equation}
Figure \ref{fig:GRBtrue} shows the intrinsic GRB rate
\begin{equation}
\frac{{\rm d}N_{\rm GRB}}{{\rm d}z} = \Psi_{\rm GRB}(z)\frac{\Delta t_{\rm obs}}{1+z}
\frac{{\rm d}V}{{\rm d}z}.
\label{dngrbtrue}
\end{equation}
In this plot, we have not considered observational effects such as 
beaming and instrument sensitivity; namely, we  set $\Omega_{\rm obs} = 4\pi, \eta_{\rm beam} =1$, 
 and $L_{lim}(z) = 0$ in Eq. (\ref{psigrb}).  
We show the GRB rate for our choice of two different IMFs. 
Interestingly, Fig. \ref{fig:GRBtrue} shows that the results 
depend only weakly on the choice of IMF. 

Figure \ref{fig:GRBtrueupp} shows the most  optimistic case,  
assuming a high star formation efficiency $f_{*} = 0.1$ for Pop~III.1, 
$f_{*} = 0.01$ for Pop~III.2, an inefficient  
 chemical enrichment, 
$v_{\rm wind} = 50$ km/s, $f_{GRB} = 0.1$, 
and a Gaussian IMF for both Pop~III.1 and Pop~III.2 stars.  
We note that constraints on these quantities will be useful 
for placing upper limits on the GRB observed rate.

\subsection{Radio afterglows}

Follow-up observations of high-redshift GRBs can
be done by observing their afterglows,  especially in radio band \citep{ioka2005,inoue2007}.  
We calculated the radio afterglow light curves for Pop~III GRBs  
following  the standard 
prescription from  \citet{sari1998,sari1999} and \citet{meszaros2006}. 
The afterglow light curve at the time $t_d$ is given by 
the shock radius $r_{\rm d}$ and the Lorentz factor $\gamma_d$.
These two quantities are related by $E_{\rm iso} \sim 4\pi r^3_{\rm d}
\gamma_{\rm d}^2 n m_{\rm p}c^2$ and $r_{\rm d} \sim c\gamma^2_{\rm d}t_{\rm d}$, 
where $n$ is the medium density and $m_{\rm p}$  the proton mass.  
The true energy is given by $E_{\rm true} = \theta^2E_{\rm iso}/2$, 
where $\theta$ is the   half opening  angle of the shock.   
The spectrum consists of power-law segments linked by critical 
break frequencies. 
These are $\nu_{\rm a}$ (the self absorption frequency), 
$\nu_{\rm m}$ (the peak of injection frequency),  
and $\nu_{\rm c}$ (the cooling frequency),  given by
\begin{eqnarray}
\nu_{\rm m} &\propto& ~ (1+z)^{1/2} g(p)^2
\epsilon_e^2 \epsilon_B^{1/2} E_{\rm iso}^{1/2} t_d^{-3/2}, \label{eq:numt}\nonumber\\
\nu_{\rm c} &\propto& ~ (1+z)^{-1/2} \epsilon_B^{-3/2}
n^{-1} E_{\rm iso}^{-1/2} t_d^{-1/2}, \label{eq:nuct}\nonumber\\
\nu_{\rm a} &\propto& ~ (1+z)^{-1} \epsilon_e^{-1}
\epsilon_B^{1/5} n^{3/5} E_{\rm iso}^{1/5}, \label{eq:nuat}\nonumber\\
F_{\nu,{\rm max}} &\propto& ~ (1+z) \epsilon_B^{1/2} n^{1/2} E_{\rm iso}
d_{L}^{-2},~ \label{eq:Fnumaxt}
\end{eqnarray}
where $g(p) = (p-2)/(p-1)$  is a function   of energy spectrum index 
of electrons $(N(\gamma_e)d\gamma_e\propto \gamma_e^{-p}d\gamma_e$, where $\gamma_{e}$ is the electron Lorentz factor),  and    
$\epsilon_{e}$ and $\epsilon_B$ are the efficiency factors \citep{meszaros2006}. 
There are two types of spectra. 
If $\nu_{\rm m} < \nu_{\rm c}$, we call it the
{\it slow cooling case}. The flux at the observer, $F_\nu$, is given by
\begin{equation}
\label{spectrumslow}
F_\nu=\left\{ \begin{array}{ll}(\nu_{\rm a} / \nu_{\rm m} )^{1/3}(\nu/\nu_{\rm a})^2 F_{\nu,{\rm max}}, &
\nu_{\rm a}>\nu, \\ 
( \nu / \nu_{\rm m})^{1/3} F_{\nu,{\rm max}}, &
\nu_{\rm m}>\nu>\nu_{\rm a}, \\
( \nu / \nu_{\rm m} )^{-(p-1)/2} F_{\nu,{\rm max}}, &
\nu_{\rm c}>\nu>\nu_{\rm m}, \\ 
( \nu_{\rm c} / \nu_{\rm m} )^{-(p-1)/2} ( \nu / \nu_{\rm c})^{-p/2}
F_{\nu,{\rm max}}, & \nu>\nu_{\rm c}. \end{array}\right.
\end{equation}
where 
$F_{\nu,{\rm max}}$ is the observed peak flux
at distance $d_{L}$ from the source.

For $\nu_{\rm m}>\nu_{\rm c}$, called the fast cooling case, the spectrum is

\begin{equation}
\label{spectrumfast}
F_\nu=\left\{ \begin{array}{ll}(\nu_{\rm a} / \nu_{\rm c} )^{1/3}(\nu/\nu_{\rm a})^2 F_{\nu,{\rm max}}, &
\nu_{\rm a}>\nu, \\ 
( \nu / \nu_{\rm c})^{1/3} F_{\nu,{\rm max}}, &
\nu_{\rm c}>\nu>\nu_{\rm a}, \\
( \nu / \nu_{\rm c} )^{-1/2} F_{\nu,{\rm max}}, &
\nu_{\rm m}>\nu>\nu_{\rm c}, \\ 
( \nu_{\rm m} / \nu_{\rm c} )^{-1/2} ( \nu / \nu_{\rm m})^{-p/2}
F_{\nu,{\rm max}}, & \nu>\nu_{\rm m}. \end{array}\right.
\end{equation}

As the GRB jet sweeps the interstellar medium, the Lorentz factor of the jet is
decelerated. When the Lorentz factor drops below $\theta^{-1}$, 
the jet starts to expand sideways and becomes detectable by the off-axis observers. 
These afterglows are not associated with the prompt GRB emission. 
Such {\it orphan} afterglows are a natural consequence of the existence of GRB's jets.  
Radio transient sources probe the high-energy population of the Universe 
and can provide further constraints on the intrinsic rate of GRBs.  
Even if the prompt emission is  highly collimated,  
the Lorentz factor drops $\gamma_{\rm d} < \theta^{-1}$ around the time
\begin{equation}
t_{\theta} \sim 2.14  \left(\frac{E_{\rm iso}}{5\times 10^{54}}\right)^{1/3}\left(\frac{\theta}{0.1}\right)^{8/3}n^{-1/3}(1+z)~ \rm days, 
\end{equation}
and  the jet starts to expand sideways. 
 Finally the shock velocity becomes  nonrelativistic around the time
\begin{equation}
t_{\rm NR} \sim 1.85\times 10^2 \left(\frac{E_{\rm iso}}{5\times 10^{54}}\right)^{1/3}\left(\frac{\theta}{0.1}\right)^{2/3}n^{-1/3}(1+z)~ \rm days, 
\end{equation}
 \citep{ioka2005}.  
After  time $t_{\theta}$,  the temporal dependence of the critical 
break frequencies should be replaced by  
$\nu_{\rm c} \propto t^{0}, \nu_{\rm m} \propto t^{-2},\nu_{\rm a} \propto t^{-1/5}$,  and $F_{\nu,{\rm max}} \propto t^{-1}$ \citep{sari1999}. We also used the same evolution in the nonrelativistic phase for simplicity, which underestimates the afterglow flux after $t_{\rm NR}$.

Figure \ref{fig:GRBafterglow} shows the light curves 
for a typical GRB from Pop~III.2 stars assuming 
an isotropic  kinetic energy $E_{\rm iso}\sim 10^{54} erg$ 
(in proportion to the progenitor mass)  as a lower limit. Pop~III.1 afterglows are expected to be brighter. 
Consistently with previous works, we conclude that it is possible 
to observe the GRB radio afterglows with 
ALMA, LOFAR, EVLA, and   SKA.

\subsection{Upper limits from radio transient survey}

\begin{figure}
\includegraphics[width=0.9\columnwidth]{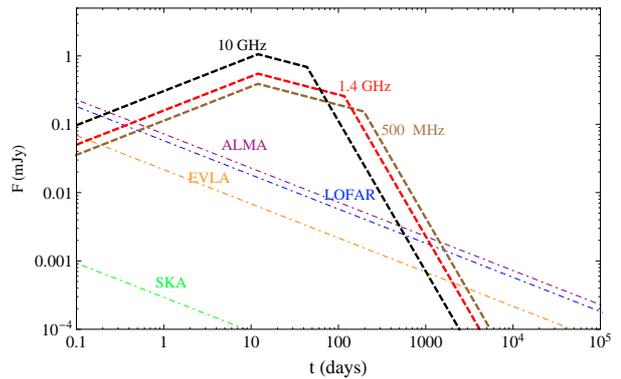}
\caption{The theoretical light curve of radio afterglow 
of a typical Pop~III.2 GRB at $z\sim 10$. 
We show the evolution of afterglow flux $F(\rm mJy)$ as a function of time $t$ (days) 
for typical parameters:  isotropic kinetic energy  $E_{\rm iso} =  10^{54}~{\rm erg}$, 
electron spectral index $p=2.5$, plasma parameters $\epsilon_{\rm e} = 0.1$,  
$\epsilon_{\rm B} = 0.01$,  
initial Lorentz factor $\gamma_{\rm d} = 200$, 
interstellar medium density  $n = 1~ {\rm cm}^{-3}$, 
for the range of frequencies: 500 MHz (dashed brown line), 
1.4 GHz (dashed red line), 
10 GHz (dashed black line),
in comparison with flux sensitivity $F_{\nu}^{\rm sen}$ 
as a function of integration time, $t_{\rm int} (\rm days)$  
for SKA (dot-dashed green line),  EVLA (dot-dashed orange line), 
LOFAR (dot-dashed blue line) and ALMA (dot-dashed purple line).}
\label{fig:GRBafterglow}
\end{figure}

In this section, we derive upper limits 
on the intrinsic GRB rate (including the off-axis GRB) 
using $\sim 1$ year timescale radio variability surveys. 
There are several radio transient surveys completed so far.
\citet{bower2007} used 22 years of archival data from VLA to 
put an upper limit of $\sim 6 ~\rm deg^{-2}$ for  1-year variability 
transients above 90 $\mu$Jy, which is equivalent to $ \lesssim 2.4 \times 10^5$ 
for the whole sky.  \citet{gal2006} used FIRST\footnote{http://sundog.stsci.edu/} 
and NVSS\footnote{http://www.cv.nrao.edu/nvss/} radio catalogs 
to place an upper limit of $\sim 70$ radio orphan afterglows above 6 mJy  
in the 1.4 GHz band over the entire sky. 
This suggests less than  $3 \times 10^4$ sources above 0.3 mJy on the sky, 
because the number of sources is expected to be proportional 
to flux limit $F_{\rm lim}^{-3/2}$ 
(assuming Euclidian space and no source evolution) \citep{gal2006}. 
From Fig. \ref{fig:GRBafterglow}, a typical GRB's radio afterglow with isotropic 
kinetic energy $E_{\rm iso} \sim 10^{54} ~ergs$  
stays above 0.3 mJy over $\sim 10^{2}$ days.  
 
\begin{figure}
\includegraphics[width=0.9\columnwidth]{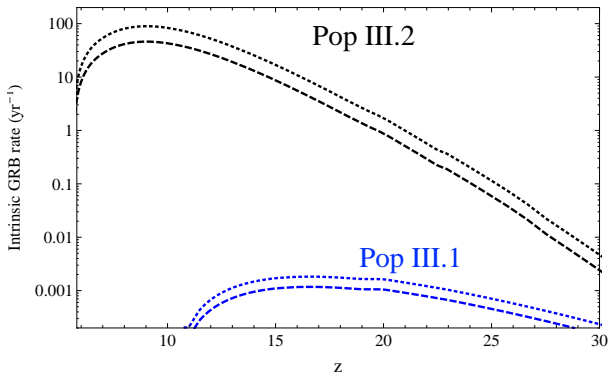}
\caption{The intrinsic GRB rate  ${\rm d}N_{\rm GRB}/{\rm d}z$.  The number of (on-axis + off-axis) GRBs per year on the sky
in Eq. (\ref{dngrbtrue}),  as a function of redshift. We set $f_{*} = 0.001$, $f_{GRB} = 0.01$ and $v_{\rm wind} = 100 {\rm km/s}$ for this plot. 
Salpeter IMF,  dashed black line,  Gaussian IMF, dotted black line, for Pop~III.2; and Salpeter IMF,  dashed blue line,  Gaussian IMF,  dotted blue line, for  Pop~III.1.}
\label{fig:GRBtrue}
\end{figure}

\begin{figure}
\includegraphics[width=0.9\columnwidth]{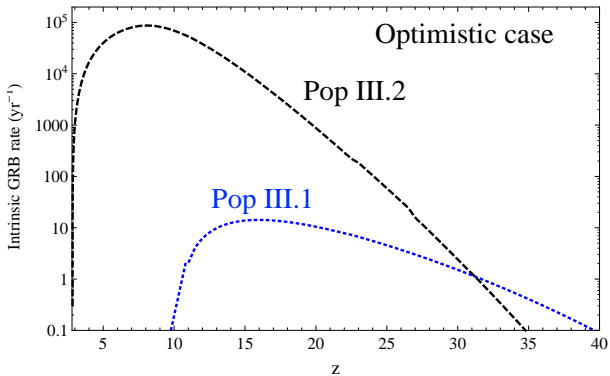}
\caption{The intrinsic GRB rate  ${\rm d}N_{\rm GRB}/{\rm d}z$.  The number of (on-axis + off-axis) GRBs per year on the sky
in Eq. (\ref{dngrbtrue}), 
as a function of redshift
for our optimistic model. We assume a high star formation 
efficiency; $f_{*}= 0.1$ for Pop ~III.1;  and $f_{*} = 0.01$ for Pop III.2;  slow chemical enrichment, 
$v_{\rm wind} = 50 {\rm km/s}$; high GRB formation efficiency, $f_{GRB}= 0.1$; and a Gaussian IMF;   
for both Pop~III.2,  dashed black line;
and Pop~III.1,  dotted blue line.}
\label{fig:GRBtrueupp}
\end{figure}

 By combining the results shown in Figs. \ref{fig:GRBtrue} and \ref{fig:GRBtrueupp}, we expect
$\sim 30-3\times10^{5}$ sources ($10^2-10^{6}$ events per year $\times 10^2$ days)
 above $\sim 0.3$ mJy. (We integrate the event rate over redshift.)  As a consequence, the most optimistic case  for Pop III.2 
should  already be ruled out marginally  by   the current observations of radio transient sources, 
if their luminosity function follows the one assumed in the present paper.
Only more conservative models are then viable.  
Radio transient surveys are not yet able to set upper limits on the Pop III.1 GRB rate.   
The above conclusion is model dependent, because the afterglow flux depends 
on the still uncertain quantities,  such as the isotropic energy $E_{\rm iso}$ and the 
ambient density $n$. If the circumburst density is higher than usual, the constraints 
from the radio transient surveys would be even stronger.  
Also the GRB formation efficiency and the beaming factor are not known
accurately, 
which can affect both the intrinsic and observed rate more than one order of magnitude. 

\begin{figure}
\includegraphics[width=0.9\columnwidth]{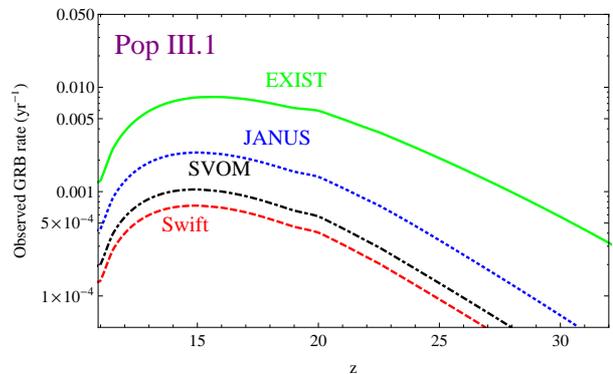}
\caption{
Predicted Pop~III.1observed GRB rate. Those 
observed by \textit{Swift}, dashed red line; SVOM, dot-dashed black line;
JANUS, dotted blue line; and EXIST, green line. We adopt a GRB rate model that is consistent with the current upper 
limits from the radio transients; Gaussian IMF, $v_{\rm wind} = 50 {\rm km/s}$, $f_{*}=0.1$, $f_{GRB} = 0.1$.} 
\label{fig:GRB2}
\end{figure}

\begin{figure}
\includegraphics[width=0.9\columnwidth]{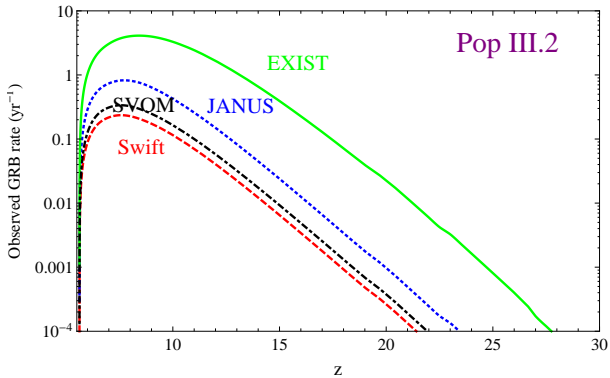}
\caption{
Predicted Pop~III.2 observed GRB rate. Those observed by \textit{Swift}, dashed red line;   SVOM, dot-dashed black line;
JANUS, dotted blue line;  and EXIST, green line;  for our model with
Salpeter IMF, $v_{\rm wind} = 100 {\rm km/s}$, $f_{*}=0.01$, $f_{GRB} = 0.01$.}
\label{fig:GRB3}
\end{figure}

In Figs. \ref{fig:GRB2}-\ref{fig:GRB3}, we show the predicted observable GRB rate 
$dN_{GRB}^{obs}/dz$ in Eq. (\ref{dngrbobs}) for Pop~III.1 
and III.2 detectable by the \textit{Swift}, SVOM,  JANUS,  and EXIST missions.  
The  results shown are still within the bounds of available upper limits from 
the radio transient surveys.   
Overall, it is more likely to observe Pop~III.2 GRBs than Pop~III.1,
but the predicted rate strongly depends on the IGM metallicity evolution,  
the  star formation efficiency and GRB formation efficiency. The dependence on the IMF is relatively small.  

\begin{figure}
\includegraphics[width=0.9\columnwidth]{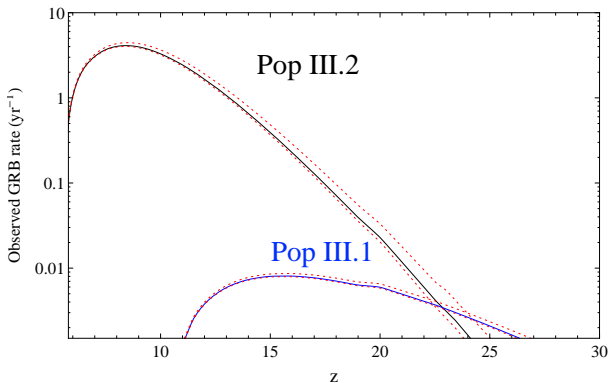}
\caption{
Predicted maximum  GRB  rates observed by EXIST.  We adopted Salpeter IMF, $v_{\rm wind} = 100 {\rm km/s}$, $f_{GRB} = 0.01$,   $f_{*}= 0.01$;  for Pop~III.2 GRBs, dashed black line; and Gaussian IMF, $v_{\rm wind} = 50 {\rm km/s}$, $f_{*}=0.1$, $f_{GRB} = 0.1$;  for Pop~III.1 GRBs, dotted blue line.  Dotted red lines represent the same with  LF's  bright end power law index 1.7 and 0.8. 
}
\label{fig:GRB5}
\end{figure}

Figure \ref{fig:GRB5} shows the GRB rate expected for EXIST observations. 
Because the power index of the LF is uncertain at the bright end, we added 
two lines to show the resulting uncertainty in our prediction.   
We use the maximum rate,  which is within the constraints by the current observations 
of radio transients. 
We expect to observe  $N \sim 20$  GRBs per year at $z > 6$ for Pop III.2 
and $N \sim 0.08$  per year for Pop III.1 at $z > 10$ with the future EXIST satellite at a maximum. 
Our optimist case  predicts  a near-future detection of Pop~III.2 GRB by  \textit{Swift}, and the nondetection so far could suggest a further upper limit or difference between the Pop III and present-day GRB spectrum.

\section{Conclusion and discussion}
\label{sec:conclusions}

There are still no direct observations of Population III stars,   
despite much recent development in theoretical studies
on the formation of the early generation stars.
In this paper,  we follow a recent suggestion that massive Pop~III stars 
could trigger collapsar gamma-ray bursts. 
Observations of such energetic GRBs at very high redshifts 
will be a unique probe of the high-redshift Universe.  
 With a semi-analytical approach we estimated the star formation rate 
for Pop III.1 and III.2 stars including all relevant feedback effects:  
photo-dissociation, reionization, and metal enrichment.

Using  radio transient sources we are able 
to derive constraints on the intrinsic rate of GRBs.  
We estimated the predicted GRB rate for both Pop~III.1 
and Pop~III.2 stars, and argued that the latter 
is more likely to be observed with future experiments.  
We expect to observe maximum of $N \lesssim 20$ GRBs per year integrated over 
at $z > 6$ for Pop III.2 and $N \lesssim 0.08$  per year integrated over at $z > 10$
for Pop III.1  with EXIST. 

We also expect a larger number of radio afterglows than X-ray 
prompt emission because the radio afterglow is long-lived, 
for $\sim 10^2$ days above $\sim 0.3$ mJy  from Fig. \ref{fig:GRBafterglow}. 
Combining with the intrinsic  GRB rate and constraints from radio 
transients, we expect roughly $\sim 10-10^4$ radio afterglows above 
$ \gtrsim 0.3$ mJy already on the sky. They are indeed detectable by ALMA, EVLA, LOFAR,  and SKA,
and  have even been detected already  by $\sim 1$ yr-timescale variability surveys.
We showed that using a semi-analytical approach combined with the 
current surveys,  such as NVSS and FIRST,  
we are already  able to constrain the Pop III.2 GRB event rate.

 Finally, it is important to note that our knowledge of the first stars and GRBs is still 
limited, and there are uncertainties in their properties, most
significantly  in their characteristic mass.
Recently, \citet{clark2011} and \citet{greif2011} have performed cosmological simulations
using a sink particle technique to follow the evolution of a primordial
protostellar accretion disk. They find that the disk gravitationally
fragments to yield multiple protostellar seeds. Although the final mass distribution
of the formed stars is still uncertain, formation of multiple systems,
especially massive binary Pop~III stars, would increase the rate of 
high-redshift GRBs \citep[e.g.,][and references therein]{bromm2006,fryer2005}. 
 If the GRB fraction per collapse $f_{\rm GRB}$ 
in Eq. (\ref{etagrb}) is
much larger than the current one, say $f_{\rm GRB} \sim 1$,
the Pop III.1 GRBs might also become detectable
with the radio telescopes ($\sim 300$ afterglows  above $\sim 0.3$ mJy on the sky)
and the X-ray satellites ($\sim 1$ event per year for EXIST)
in the future.

\begin{acknowledgements}
R.S.S. thanks the Brazilian agency CNPq (200297/2010-4)  
for financial support. This work was supported by 
World Premier International Research Center Initiative 
(WPI Initiative), MEXT, Japan.
We thank Emille Ishida, Andrea Ferrara, Kenichi Nomoto, Jarrett  Johnson,
and Yudai Suwa for fruitful discussion and suggestions.  
NY acknowledges the financial support by 
the Grants-in-Aid for Young Scientists (S) 20674003
by the Japan Society for the Promotion of Science. KI acknowledges the financial support  by KAKENHI  21684014, 19047004, 22244019,  22244030. We  thank the anonymous referee for their very careful reading of the paper and for several suggestions which allowed us to  improve the current work. We also thank the language editor J. Adams for his carefully revision. 
\end{acknowledgements}

\bibliographystyle{aa}

\end{document}